%
%
%
\newcount\eqnumber
\eqnumber=1
\newcount\eqnumbA
\eqnumbA=1
\newcount\fignumber
\fignumber=1
\def\neqn{{\rm(\the\eqnumber)}\global\advance\eqnumber by 1}
\def\neqnA{{\rm(A\the\eqnumbA)}\global\advance\eqnumbA by 1}
\def\nfig{\global\advance\fignumber by 1}
\def\eqnam#1#2{\immediate\write1{\xdef#2{(\the\eqnumber}}
\xdef#1{(\the\eqnumber}}
\def\eqnamA#1#2{\immediate\write1{\xdef#2{(A\the\eqnumbA}}
\xdef#1{(A\the\eqnumbA}}
\def\fignam#1#2{\immediate\write1{\xdef\ #2{\the\fignumber}}
\xdef#1{\the\fignumber}}
\def\spose#1{\hbox to 0pt{#1\hss}}
\def\tophat{\spose{\raise 0.5ex\hbox{\hskip4.0pt$\widehat{}$}}}
\def\topTwi{\spose{\raise 0.5ex\hbox{\hskip4.0pt$\widetilde{}$}}}
\def\toptwi{\spose{\hbox{\hskip2.0pt$\widetilde{}$}}}
\input mn.tex
\input epsf.tex
\hyphenation{Pij-pers}
%
%

\loadboldmathnames

\begintopmatter

\title{The determination of time lags using SOLA}
\author{ Frank~P.~Pijpers}
\affiliation{Teoretisk Astrofysik Center, Danmarks Grundforskningsfond,
     Institut for Fysik og Astronomi, Aarhus Universitet,
     DK-8000~Aarhus~C, Denmark}
\shortauthor{F.P. Pijpers }
\shorttitle{SOLA time-lag determination}
\acceptedline{}
\abstract{A common problem in astronomy is the determination of the time
shift between two otherwise identical time series of measured flux
from a variable source, in short the determination of a time lag. Two
examples of where this problem occurs are in the determination of the
Hubble constant from multiple images of gravitationally lensed variable
quasars and also in the determination of distances to OH/IR stars. It
is shown here that this problem can be seen as a restricted inversion
problem. It therefore can be solved using the subtractive optimally
localized averages (SOLA) method for inversion which has been described
elsewhere (Pijpers \& Thompson 1992, 1994~; Pijpers \& Wanders 1994).
}
\maketitle

\section{Inverse problems}

The problem of determining a time lag between two time series of
fluxes is very similar to the problem of reverberation mapping of active
galactic nuclei. The difference lies primarily in what is known about the
so-called transfer function (cf. Blandford \& McKee 1982~; Peterson 1993).
Essentially the problem of reverberation mapping comes from a view of
an AGN as gas clouds surrounding a variable continuum source. The
gas clouds re-emit the radiation absorbed from the continuum source in
spectral lines so that the time lag between the variation of the line
emission and the continuum emission is a measure of the difference in
path length to the observer and hence of the distance from the central
source of the emitting gas clouds. The transfer function is thus related
to the distribution of clouds around the nucleus.
Mathematically, the concept of reverberation mapping leads to the integral
equation
\eqnam\RevMap{RevMap}
$$
L (t)=\int{\rm d}\tau\ \Psi(\tau) C(t - \tau).
\eqno\neqn
$$
Here $L$ and $C$ are the (velocity integrated) line flux of a broad
line in the AGN spectrum, and the continuum flux respectively.
Hence the problem is reduced to the inversion of the integral equation
to obtain the transfer function $\Psi$.
If in equation \RevMap) for $\Psi(\tau)$ a Dirac delta function
is substituted, $\Psi(\tau) = \delta (\tau -t_0)$, equation \RevMap)
reduces to $L(t) = C(t-t_0)$.
Conversely if two light curves are related by a time delay $t_0$ this is
equivalent to equation \RevMap) with a transfer function $\Psi(\tau) =
\delta (\tau - t_0)$. In this paper the transfer function is therefore
assumed to be essentially zero everywhere except at the unknown time-lag
$t_0$ : $\Psi ( \tau) = I \delta (\tau - t_0)$ where $\delta$ is the
Dirac delta function.

Contrary to the problem of reverberation mapping where there is an
assumed causal relationship between the variations in continuum and
line flux, in the problem of time lag determinations the light curves
are not distinguishable as driving or responding time series.
The inversion method should reflect this lack of knowledge in a symmetric
treatment of the two time series.
The notation of equation \RevMap) is slightly modified to
express this~:
\eqnam\PhaseL{PhaseL}
$$
\eqalign{
F^{(b)} (t)= &\int\limits_{-\tau_{\rm max}}^{\tau_{\rm max}}{\rm d}
\tau\ I\delta(\tau - t_0 ) F^{(a)} (t -\tau ) \cr
F^{(a)} (t)= &\int\limits_{-\tau_{\rm max}}^{\tau_{\rm max}}{\rm d}
\tau\ {1\over I} \delta(\tau + t_0 )
F^{(b)} (t -\tau ) \cr}
\eqno\neqn
$$
The two equations both express that there is a time lag $t_0$ between the
two time series $F^{(a)}$ and $F^{(b)}$ and therefore seem redundant.
However both need to be used to ensure a symmetric treatment of the two
time series in the algorithm. $I$ is a constant to allow for a constant
non-unity ratio between the two time series. Note that $I$ can have any
value larger than $0$ and that the time-lag $t_0$ can be either positive
or negative, since it is not known a-priori which light curve is leading
and which is lagging. Just as in the application of the subtractive
optimally localized averages method (SOLA) to reverberation mapping
(cf. Pijpers \& Wanders, 1994~: hereafter PW) the limits of the
integration are set to a finite value. The reason for this is that for
any measured time series its total extent is finite and therefore the
range over which a time lag can be determined is limited to values within
this range.

Equations \RevMap) and \PhaseL) are idealized in the sense that
finite sampling rates of the two light curves and finite measurement
errors are not yet explicitly accounted for. This is done in the following
section. Another problem can be that one or both of the light curves
are contaminated by a foreground or background source. This can be dealt
with, however, and the procedure is described in the appendix to this paper.

\section{A brief description of SOLA}

The strategy of the SOLA method in general is to find a set of linear
coefficients which, when combined with the data, produce the value of
the unknown convolved function under the integral sign for given value(s)
of the time lag. In order to do this the time series under the integral
sign is interpolated. To each measurement of the series outside of the
integral sign is assigned a partial time series which is a section of
the time series under the integral sign.
As discussed by Pijpers and Wanders (PW) this means that to each
measurement $L(t_i)$ the partial time series consisting of the $i^{th}$
measurement $C(t_i)$ and all $n$ previous ones, $[C(t_{i-n}), C(t_i)]$, is
assigned. These partial time series $[C(t_{i-n}), C(t_i)]$ for all $i$
form the set of base functions for the SOLA method when it is applied
to equation \RevMap).

\fignam{\Figone}{Figone}
\beginfigure{1}
\epsfxsize=7.5cm
\epsfbox{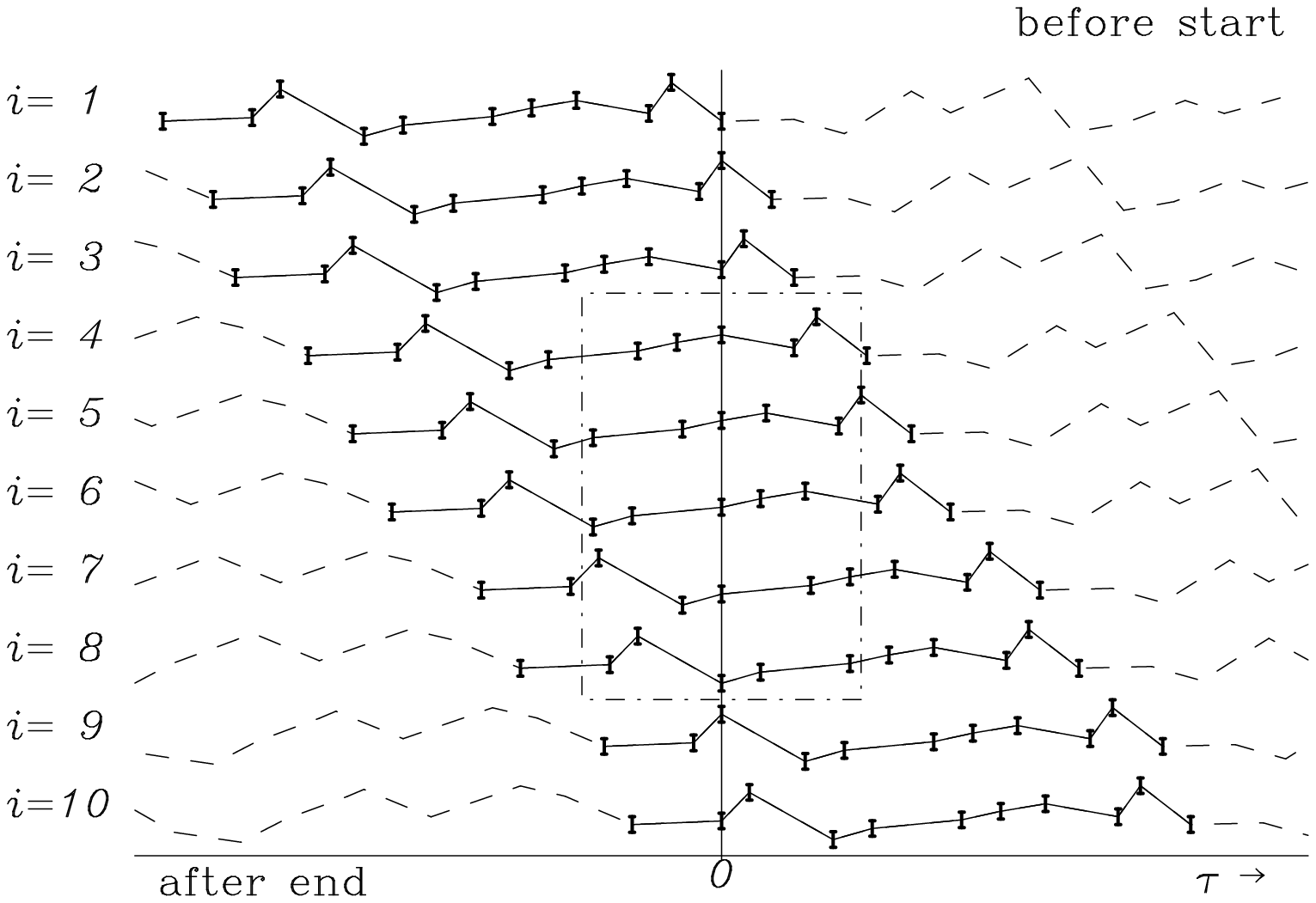}
\caption{{\bf Figure \Figone.} Example of two time series with 11
measurements.
The entire time series under the integral sign is
re-plotted 10 times with an arbitrary vertical offset for each of
the measurements of the time series outside the integral sign.
The horizontal scale is time delay which means that later measurements
fall further to the left.
The solid part is the part of the time series that is actually
measured, the dashed parts fall before the first or after the last of
the measurements.}
\endfigure\nfig

These base functions are used to build a localized averaging kernel using a
set of linear coefficients $\{ c_i \}$ determined for this purpose. In other
words if $P_i^{(n)}$ denotes the (interpolated) partial time series
$[C(t_i), C(t_{i+n})]$, which is normalized to have an integral of
$1$, and $F_i = F(t_i)$ corrected for the normalization of the
$P_i^{(n)}$, then
\eqnam\Summs{Summs}
$$
\eqalign{
\sum\limits_{i} c_i P_i^{(n)} &= {\cal K} (\tau, \tau_0) \cr
\sum\limits_{i} c_i F_i &= \psi (\tau_0) \cr
}
\eqno\neqn$$
where ${\cal K}$ is an integration kernel that is localized around
$\tau_0$ and $\psi (\tau_0)$ is the associated localized average of
$\Psi$ around $\tau_0$. The details of this procedure can be found
in the papers describing the SOLA method (Pijpers \& Thompson 1992, 1994
hereafter PT1 and PT2) and its application to reverberation mapping (PW).
In many respects the same technique is followed here.
Figure 1 is included to assist in the visualization of this procedure,
where the entire time series is plotted as a function of time delay.
The entire time series under the integral sign is
re-plotted 10 times with an arbitrary vertical offset for each of
the measurements of the time series outside the integral sign.
The integration limits $-\tau_{\rm max}$ and $\tau_{\rm max}$ are shown
as the vertical dash-dotted lines.
It is clear that if $\tau_{\rm max}$ is set to a value larger
than or equal to half the length of the time series no base function
will be defined over the entire integration interval, which renders
the summation in equation \Summs) meaningless. Limiting the range
of integration to smaller values, and therefore the possible range over
which a time lag can be determined, has the result that sections of
the time series can be used as base functions. These are the sections
between the vertical dash dotted lines in figure 1. The first few and
last few measurements then have partial time series associated with them
that still do not cover the entire range $[-\tau_{\rm max},
\tau_{\rm max}]$ which means that these must be excluded from the analysis.
The remaining points have associated partial time series that are defined
within the dash-dotted window in figure 1. The total range of integration
$[-\tau_{\rm max}, \tau_{\rm max}]$ must be strictly smaller than the
total length of the measured time series. $\tau_{\rm max}$ is a free
parameter of the method and since $t_0$ is unknown it may well be
necessary to explore a range of values for $\tau_{\rm max}$ and re-do the
inversion for each value.
Note that one can increase $\tau_{\rm max}$ at the cost of having less base
functions, i.e. decrease the height and increase the width of the
window, and vice versa. To construct a good approximation to the target
form for the integration kernel from the base functions it is desirable
to have as many base functions as possible, which implies a small
$\tau_{\rm max}$. However, to obtain a reliable estimate of the time lag
it is necessary to ensure that $\tau_{\rm max}$ is larger that the
expected $t_0$.

The time series is interpolated using Savitzky-Golay interpolation
(cf. Press et al., 1992). In a Savitzky-Golay filter a polynomial of
fixed degree is fitted to a moving window with a fixed number of measured
data points.
In this implementation of the SOLA algorithm the data points are chosen
symmetrically around the subinterval for which the interpolation is done.
The base functions in this implementation of the SOLA method are
subsections of the entire time series and are known to within the errors
propagated from the measured points by the polynomial fitting algorithm.

Contrary to the application
of SOLA to reverberation mapping the kernel ${\cal K}$ will not be
designed to be localized. To obtain an estimate of the parameter
$t_0$ which is the position of the peak of the very narrow transfer
function appropriate for simple time lags it is much better to
determine the first moment of the transfer function.
It is clear that if the transfer function is narrow compared to the
sampling rate or intrinsic time scale of variability of $F$ one cannot
hope to reconstruct its shape in an inversion. The width of the
reconstructed transfer function will in this case be almost identical
to the resolution given by the sampling of the
time series (cf. PW and Pijpers 1994) rather than the real width. For
simple time lag determinations this is not important, however, since only
the single unknown time lag $t_0$ needs to be determined.
To do this most conveniently one should combine the
series of partial light curves of the time series under the integral
sign into a kernel that is not the usual Gaussian but a linear
function of $\tau $. This means minimizing for two sets of coefficients
$\{ c_i^{(1a)} \}$ and $\{ c_i^{(1b)} \}$~:
\eqnam\minimize{minimize}
$$
\eqalign{
\int\limits_{-\tau_{\rm max}}^{\tau_{\rm max}}{\rm d}\tau\left[
\sum c_i^{(1a)} F^{(a)} (t_i-\tau ) \right. &\left. - {\cal T} \right]^2 \cr
&+ \mu_0 \sum c_i^{(1a)} c_j^{(1a)} E_{ij} \cr
\int\limits_{-\tau_{\rm max}}^{\tau_{\rm max}} {\rm d}\tau\left[
\sum c_i^{(1b)} F^{(b)} (t_i-\tau ) \right. &\left. - {\cal T} \right]^2 \cr
&+ \mu_0 \sum c_i^{(1b)} c_j^{(1b)} E_{ij} \cr }
\eqno\neqn
$$
In both cases the target function ${\cal T}$ for the averaging kernel
is linear~:
\eqnam\lintarfun{lintarfun}
$$
{\cal T}\ \equiv\ \tau
\eqno\neqn
$$
The factor $\mu_0$ is a free parameter
which can be used to adjust the relative weighting of the errors in
the variance-covariance matrix $E_{ij}$ and the approximation of the
target form. The use of this parameter has been described in the papers
PT1 PT2 and in PW. Its purpose is to balance a matching of the target
kernel function against magnification of the errors which are generally
opposing aims. An extra constraint is imposed on the coefficients
which is that~:
\eqnam\constraint{constraint}
$$
\eqalign{
\sum c_i^{(1a)}\ &=\ 0 \cr
\sum c_i^{(1b)}\ &=\ 0 \cr}
\eqno\neqn
$$
Imposing this constraint ensures that the integral of the averaging kernel
is equal to $0$ just as the integral of the target kernel. In this way
any influence of the even-order moments of the transfer function is
eliminated. The result of the minimization of \minimize) is that
\eqnam\conskern{conskern}
$$
\eqalign{
\sum c_i^{(1a)} F^{(a)}(t_i -\tau) &\approx \tau \cr
\sum c_i^{(1b)} F^{(b)}(t_i -\tau) &\approx \tau \cr}
\eqno\neqn
$$
The superscript for the coefficients $c_i$ identifies which of the time
series $a$ or $b$ is under the integral sign, and therefore used to build
the kernel, and the $1$ signifies that the linear target kernel \lintarfun) was
used. When \PhaseL) and \conskern) are combined the result is~:
\eqnam\steps{steps}
$$
\eqalign{
\tophat{I t_0}^{(a)}\ &\equiv\ \sum c_i^{(1a)} F_i^{(b)}\ \cr
&=\ \int {\rm d}\tau I\delta(\tau - t_0) \sum c_i^{(1a)} F^{(a)}(t_i-\tau) \cr
&\approx\ \int {\rm d}\tau I\delta(\tau - t_0) \tau \cr
&= I t_0 \cr
-{\tophat{t_0}^{(b)} \over I}\ &\equiv\ \sum c_i^{(1b)} F_i^{(a)}\ \cr
&=\ \int {\rm d}\tau {1\over I}\delta(\tau + t_0) \sum c_i^{(1b)} F^{(b)}
(t_i-\tau) \cr
&\approx\ \int {\rm d}\tau {1\over I}\delta(\tau + t_0) \tau
\cr
&= -{1\over I} t_0 \cr}
\eqno\neqn
$$
where $\tophat{t_0}$ is the estimated time lag. It is clear that the
right-hand sides do indeed give estimates $\tophat{t_0}$ of the
position $t_0$ of the delta function. The factor $I$ can be determined
independently by using the SOLA method with a kernel that is constant
over the entire range of integration as described in PW~;
${\cal T}= 1/2 \tau_{\rm max}$ instead of ${\cal T} = \tau$. (Note that
the constant is chosen to obtain normalization with integral $1$ on the
interval $[-\tau_{\rm max}, \tau_{\rm max}]$). For this target form the
associated coefficient sets are $\{ c_i^{(0a)}\}$ and $\{ c_i^{(0b)}\}$.
For the coefficients $c^{(0a)}, c^{(0b)}$ in the constraint \constraint)
the sum should be equal to $1$ instead of $0$. Note that the normalization
of the base functions is assumed to have been carried out before
the kernel is constructed in the minimization of \minimize).

The effect of data-errors in the measurements on the left-hand sides of
the equality in equations \PhaseL) is taken into account in the usual
way (cf. PW). Since the result is a linear combination of the data the
resulting error estimate can be computed trivially. The primary reason to
prefer the method described here to determining the shape of the transfer
function is that the magnification of data errors is expected to be much
smaller with the method described here.
Low order moments of $\Psi$ such as the zero order moment $I$ and
the first order moment $It_0$ can be determined with a higher accuracy
because there is a roughly inversely proportional relation between the
resolution width of the localized kernel ${\cal K}$ and the magnification
of data errors.

The effect of data errors in the measurements under the integral sign
is not quite as straightforward as for the errors outside of the
integral sign. The most important point to realize is that the kernel
that is constructed will
in general not match perfectly the target function. It is for this reason
that the third equality in the two equations \steps) is only approximate.
This means that the estimates $\tophat{t_0}$ must be corrected for such
deviations from the target form, which can be done with the help
of the constructed averaging kernel. In practice this means finding
at which $t_0$ a delta function must be placed to obtain the estimate
$\tophat{t_0}$ given the averaging kernel that was constructed. This is
straightforward since it requires only a simple function evaluation
with the constructed averaging kernel as the function.
For a perfect match between those two functions $t_0 = \tophat{t_0}$.
For real data, with only a limited number of base functions to
work with, there is always a small correction $e$ due to the deviation
of the constructed kernel from the target form. Except in those sections
where the kernels themselves are discussed, from this point the correction
$e$ is implicit in the estimates $\tophat{t_0}$~:
\eqnam\corrests{corrests}
$$
\eqalign{
\tophat{t_0}^{(a)} &= {\sum c_i^{(1a)} F_i^{(b)} \over \sum c_i^{(0a)}
F_i^{(b)} }+ e^{(a)} \cr
\tophat{t_0}^{(b)} &= {\sum c_i^{(1b)} F_i^{(a)} \over \sum c_i^{(0b)}
F_i^{(a)} }+ e^{(b)} \cr}
\eqno\neqn
$$
For a determination of the time lag that is explicitly symmetric in the
treatment of the two time series, the mean of the
time lags $T_0$ from the two alternatives is taken. Half the
difference between the two alternatives $Z$ should be equal to zero to
within the same errors as apply to $T_0$.
$$
\eqalign{
T_0 &\equiv {1\over 2}\left( \tophat{t_0}^{(a)} + \tophat{t_0}^{(b)}
\right) \cr
Z &\equiv {1\over 2} \left\vert \tophat{t_0}^{(a)} - \tophat{t_0}^{(b)}
\right\vert \cr}
\eqno\neqn
$$
where \corrests) is used to calculate the $\tophat{t_0}$.
It is immediately clear that a time lag $T_0$ thus
determined is invariant under interchange of the two time series.
$Z$ can be used as an additional tool to gauge the accuracy with which
the lag is determined in the algorithm since it should be equal to $0$
to within the errors. If it is not this can be an indication of
contamination of the time series by an extraneous source. Its value, together
with the difference between the two determinations (from interchanging the
time series) of the magnification $I$, can be used to correct for such
contamination as is demonstrated in the appendix.

\section{Artificial data}

\fignam{\simLCV}{simLCV}
\beginfigure{2}
\epsfxsize=7.5cm
\epsfbox{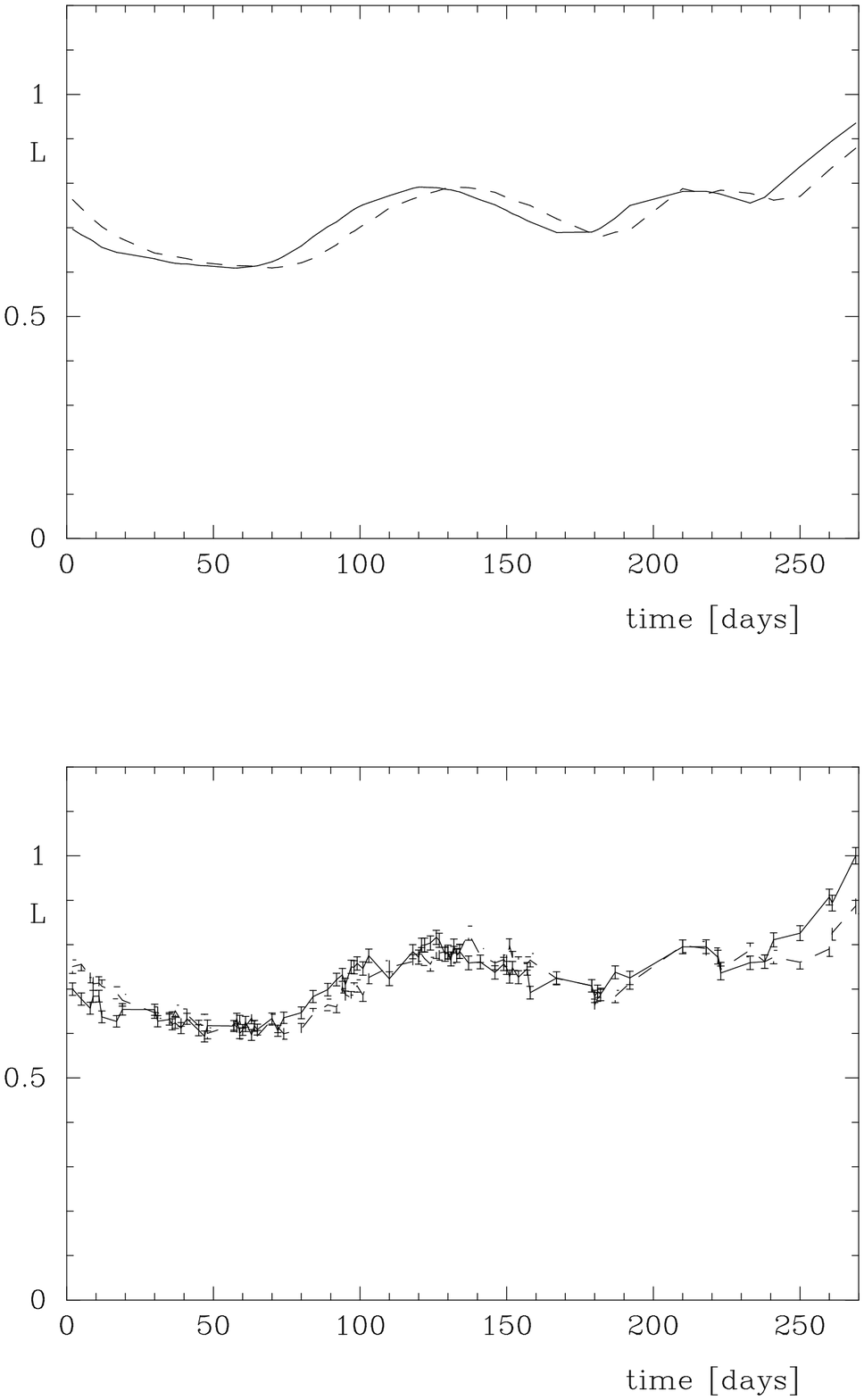}
\caption{{\bf Figure \simLCV.} The light curves that were used for the
simulations, a. without noise, b. with 2 \% noise. }
\endfigure\nfig

\fignam{\kernels}{kernels}
\beginfigure*{3}
\epsfysize=11.4cm
\epsfbox{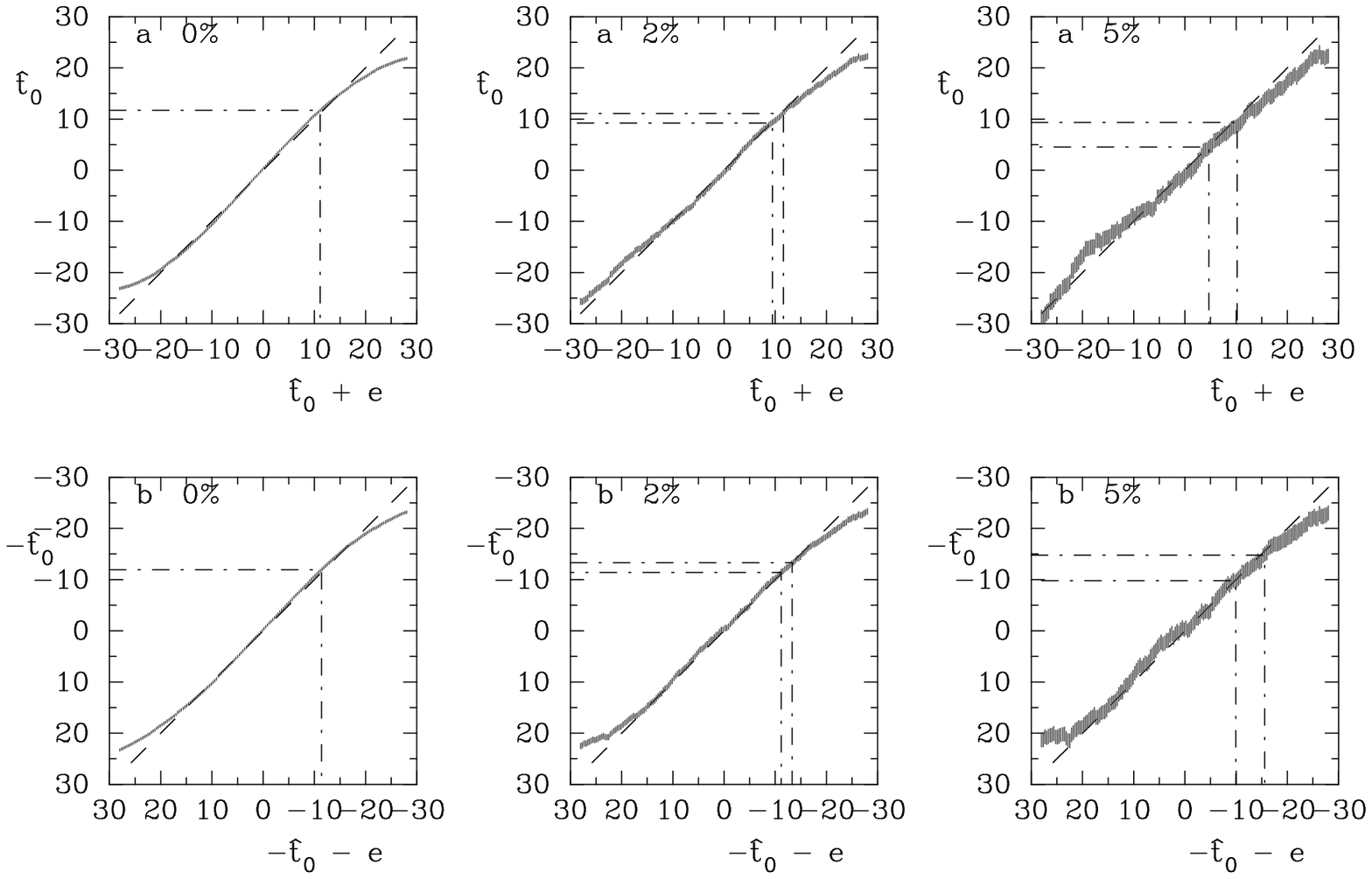}
\caption{{\bf Figure \kernels.} The constructed averaging kernels
for $\tau_{\rm max} = 28\ {\rm d}$. Left-hand panels are for the case
without noise, middle panels have 2 \% noise, right-hand panels have
5 \% noise, The grey area denotes the extent of the error bar on the
constructed kernel for the cases with noise. The dash dotted lines
denote a $2\sigma$ interval centred at $\tophat{t_0}$. Note that in the
lower panels where the role of the time series is reversed with respect
to the upper panels, the axes are also reversed to facilitate visual
comparison of the time lag $\tophat{t_0}+e$ determined for each case.}
\endfigure\nfig

To assess the influence of data errors on the algorithm and possible
systematic effects the method was tried out on a set of artificial
data. The light curves with and without errors are shown in figure
\simLCV{}. This light curve is a smoothed version of a continuum light
curve reported by Peterson et al. (1994) for the active galaxy NGC~5548.
The solid line is the original light curve, the dashed line
is the light curve after convolution with a Gaussian with a width of
$0.1\ {\rm d}$ and a central position of $11.3\ {\rm d}$. These numbers
were chosen arbitrarily
but the requirement for this method that the transfer function
be sharply peaked is satisfied. Also the central position is chosen not
to be commensurate with any sampling interval. The irregular sampling
intervals range from a minimum of $1\ {\rm d}$ to several days. The
second panel of figure \simLCV{} shows both of the light curves with
random noise added drawn from a normal distribution with a sigma of
$2\ \%$ of the flux. Apart from using error free data ($0\ \%$),
and $2\ \%$ random noise a final case with $5\ \%$ random noise added
to the `fluxes' is also considered.

It should be noted that although it is usual to measure
noise compared to the measured flux, it is actually more appropriate
to compare the noise with the amplitude of the actual variations
in the light curves. An arbitrary constant multiplication factor
between the fluxes of the measured time series contains no information
about a time lag. In the
artificial case at hand this means that the actual S/N is $\sim 25$ for
the $2\ \%$ case and $\sim 10$ for the $5\ \%$ case.

\fignam{\realcur}{realcur}
\beginfigure{4}
\epsfxsize=7.5cm
\epsfbox{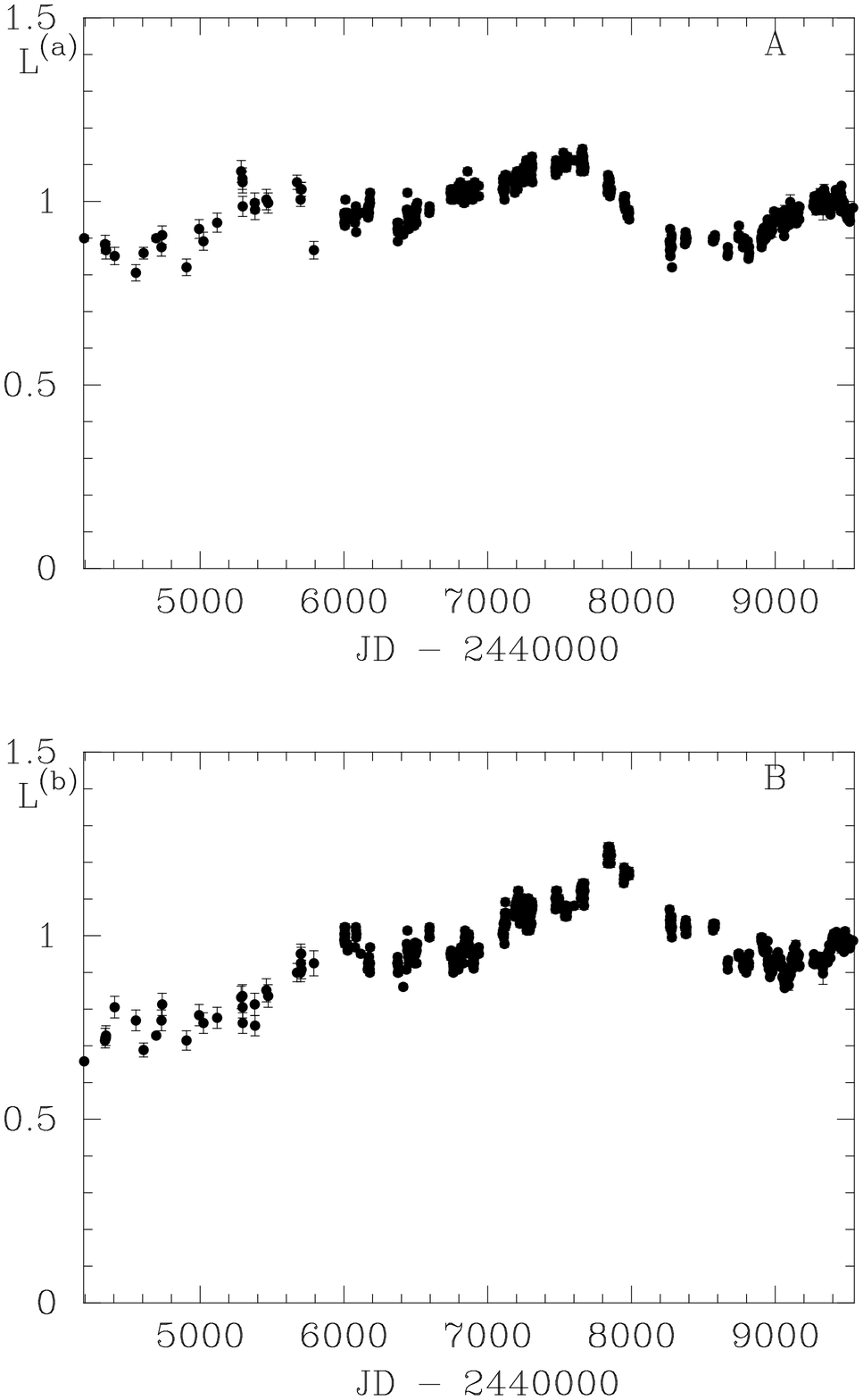}
\caption{{\bf Figure \realcur.} The optical light curves of images
A and B of the gravitationally lensed QSO 0957+561. The fluxes for the
two images are in the same units chosen such that the values are of
order unity. (Schild \& Thomson, 1995) }
\endfigure\nfig

The first reconstruction is done taking $\tau_{\rm max} = 55\ {\rm d}$
which is about $1/5$ of the total length of the time series, the second is
done taking $\tau_{\rm max} = 34\ {\rm d}$ which is about $1/8$ of the
total length of the time series, finally a $\tau_{\rm max} = 28\ {\rm d}$
which is about $1/10$ of the total length is used. These three choices
of $\tau_{\rm max}$ were applied to all three pairs of time series
contaminated by noise at $0\ \%$, $2 \%$, and $5\%$ of the fluxes $F_i$
respectively.
In each case $\tau_{\rm max}$ is the length of the shortest partial time
series $[t_i, t_{i+k}]$ that is longer than the value of $\tau_{\rm max}$
chosen as input for the algorithm.
Figure \kernels{} shows the constructed averaging kernels for
$\tau_{\rm max} = 28\ {\rm d}$. Shown in table 1 are $T_0$ and $Z$, the
final column shows the error estimates from the propagated data errors.
The error estimate is $0.0$ for the first 3 entries because for these
no noise is added to the time series. For the cases with added noise
$Z$ is within $2\sigma$ of $0$ as would be expected. The only exception
is the case for $\tau_{\rm max} = 55$ where there is a $2.5 \sigma$
departure from $0$. It is clear from the large values of $Z$ for the
choice of $\tau_{\rm max} = 55\ {\rm d}$ for the error free data and
for the data with random noise added that this value of $\tau_{\rm max}$
is still too large to obtain a reliable estimate of the time lag, although
for the cases where noise is added the deduced value of $T_0$ is within
$1 \sigma$ of the true value.
There is only marginal reduction of the error estimates when
reducing $\tau_{\rm max}$ from $34\ {\rm d}$ to $28\ {\rm d}$ and at
least in this realization of the $5\ \%$ noise the estimate of the
time lag is less accurate for the $\tau_{\rm max} = 28\ {\rm d}$ case.

\begintable{1}
\caption{{\bf Table 1.} The simulations. Error weighting $\mu_0 = 0.01$}
\halign{#\hfil&\ #\hfil&\hfil#\quad&\quad#\hfil&\quad#\hfil\cr
\noalign{\hrule\medskip}
noise& $\tau_{\rm max}$ & $T_0$ & $Z$ & $\sigma(T_0)$\cr
\noalign{\hrule\medskip}
 0\% & 55.0 & 10.2 & 1.1 & 0.0\cr
 0\% & 34.0 & 10.9\rlap9 & 0.26 & 0.0\cr
 0\% & 28.0 & 11.2\rlap7 & 0.13 & 0.0\cr
 2\% & 55.0 & 11.0 & 4.5 & 1.7\cr
 2\% & 34.0 & 11.4\rlap9 & 0.78 & 0.89\cr
 2\% & 28.0 & 11.3\rlap7 & 0.85 & 0.75\cr
 5\% & 55.0 & 9.9  & 6.5 & 4.4\cr
 5\% & 34.0 & 11.0 & 2.6 & 2.3\cr
 5\% & 28.0 & 10.1 & 2.7 & 2.0\cr
\noalign{\medskip\hrule}}
\endtable

In figure \kernels{} plots of kernels constructed from the artificial
data of figure \simLCV{} are shown. In this figure the estimates
$\tophat{t_0}$ are on the ordinate scale and the values corrected for
kernel deviations (cf. equation \corrests) ) are on the abscissa. It can
be seen that the kernel suffers somewhat from `edge
effects' where the target kernel is only poorly matched. The constructed
kernel is flatter than proportional near the edges of the integration
range. The effect this has is that at the edges of the interval
$[-\tau_{\rm max}, \tau_{\rm max}]$ the uncertainty in the time lag
determination is larger than indicated by the formal errors.
Also, if the real time-lag $t_0$ is smaller than $\sim -20\ {\rm d}$ it
will be overestimated by $\tophat{t_0}$, and if the time-lag $t_0$ is
larger than $\sim 20\ {\rm d}$ it will be underestimated by $\tophat{t_0}$.
This result indicates that not only should $\tau_{\rm max}$ be chosen
significantly less than half the total length of the measured time series,
but also significantly larger than the expected time lag.
{}From these simulations it is clear that given a total extent of the
artificial time series $T_{\rm tot} = 267\ {\rm d}$, $\tau_{\rm max}$
should be chosen $< 55\ {\rm d}$. A `safe' criterion seems to be~:
$$
\tau_{\rm max} \leq {T_{\rm tot}\over 8}
\eqno\neqn
$$
Furthermore because of inevitable `edge effects' in the kernel the true
time lag $t_0$ should be no more than $\sim 0.7 - 0.8 \times \tau_{\rm max}$.
This implies that if one wishes to measure reliably a time lag using this
method the total length of the measured time series should be at least
$10 - 12$ times the time lag $t_0$ that is expected.

\section{The sampling strategy of the time series of QSO 0957+561}

A crucial aspect of this method and most other methods in use to determine
time delays is the interpolation of the time series, which is almost always
necessary to estimate time delays. Interpolation schemes are usually
very sensitive to the sampling of the time series. Before continuing
with SOLA time delay determinations it is useful therefore to examine in
more detail the influence of the sampling on the SOLA method.

\subsection{The distribution of measurement points in time}

The largest homogeneous data set for the two optical images
of the gravitationally lensed QSO 0957+561 is that reported by
Schild \& Thomson (1995), who also have made available a master set
which combines data from other sources. The time series for the former
are shown in the two panels of figure \realcur{}. The total extent of
these time series is $5347\ {\rm d}$ and so a time lag $t_0$ of up to
$\sim 550\ {\rm d}$ can be determined with reasonable accuracy with this
method and these time series. Rather than use magnitudes the time
series are converted to (arbitrary) flux units. The reason for this is that
if there is any contamination present in the time series, due to extraneous
factors such as foreground or background sources or micro-lensing, this
is additive in flux and not in magnitude. Such contamination can be
corrected for to some extent as long as it is additive, as demonstrated
in the appendix. The flux is calculated from the magnitudes using~:
\eqnam\fluxun{fluxun}
$$
F_i = 4\ 10^5\ \exp{\left( - m_i /2.5 \right)}
\eqno\neqn
$$

Schild and Thomson (1995) note that since the more recent part
of their measured data series has sections with sampling rates that are
as high as once per day, there should be no lack of high frequency signal
in the time series with which to determine a time lag with high accuracy.
This is true in principle but it does depend to some extent on the manner
in which these sections are distributed within the overall time series.
To illustrate this point all those measurement dates were isolated from
the time series for which there are also measurements on the previous
two days and the following two days. Thus all these points are the
middle point of a section of 5 measurements done on consecutive days.
In the entire time series there are 366 such quintuplets, many of which
partially overlap because for some periods there are even more than 5
consecutive days on which measurements are available. In order
to enable using the high frequency signal in these quintuplets when
determining the time lag they must overlap with another quintuplet after
the time series is shifted by that time lag $t_0$. If the time separation
between a given pair of quintuplets is the exact time lag between the
two quasar images the measured flux in the two images should behave
exactly the same (in the absence of measurement errors), so the
quintuplets should be exactly the same. If two quintuplets do not overlap
after shifting, whatever high frequency signal they contain does not have
a measured counterpart in the other time series and is virtually useless.
The time separation between the middle points of each of these pairs is the
time lag for which that pair can be used optimally to determine whether
it is the actual time lag between the two quasar images.
The more quintuplets overlap for all possible time lags the better the
true time lag can be determined with the measured series. Of course it is
quite arbitrary to choose 5 consecutive days. One could as well take
any other number or take all weeks for which there are a certain minimum
number of measurements done. This will change some of the histograms to
be shown in this section but not the essentials of the arguments in
favour of carefully designing sampling strategies.

\fignam{\sampling}{sampling}
\beginfigure{5}
\epsfxsize=7.5cm
\epsfbox{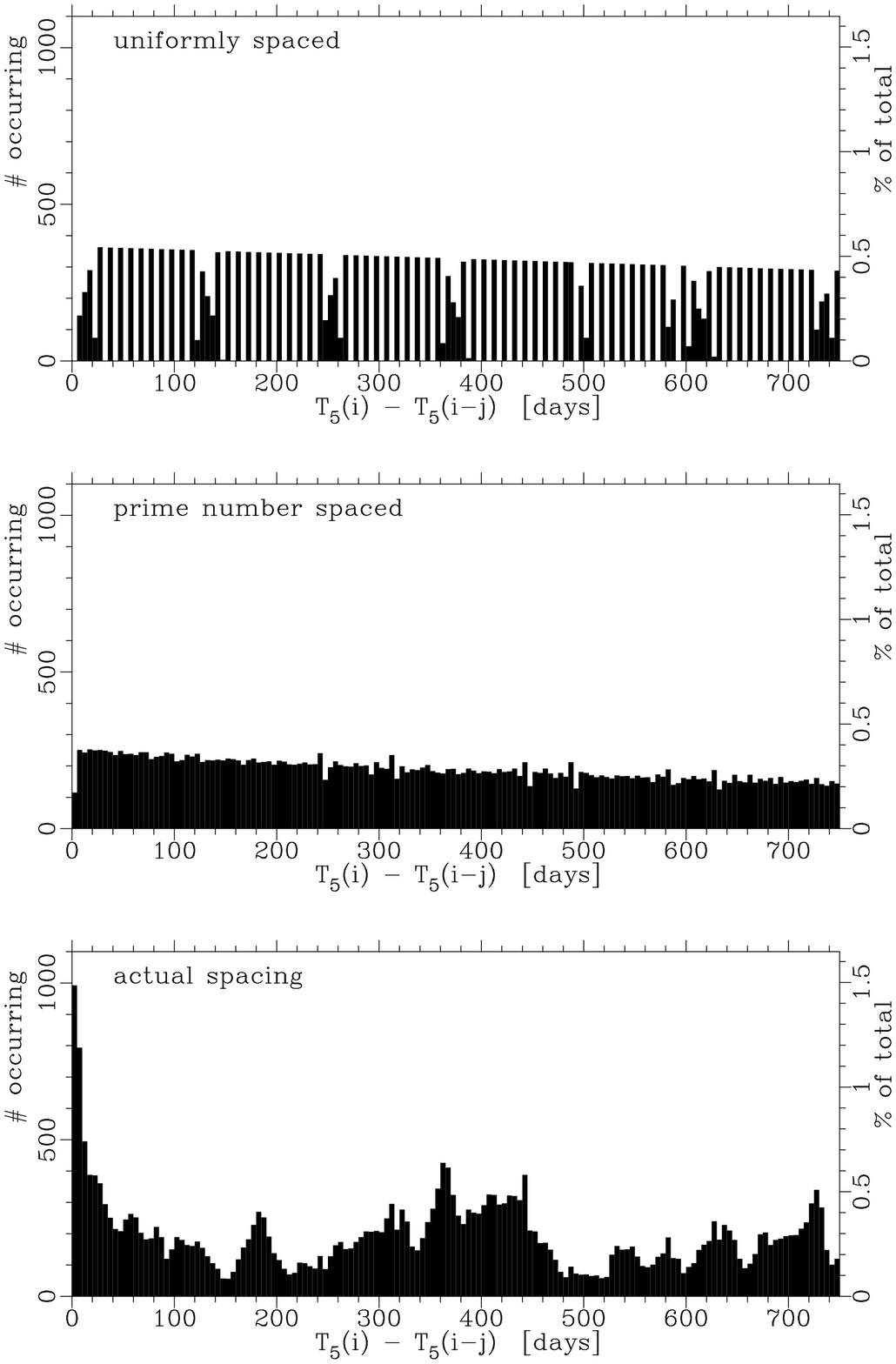}
\caption{{\bf Figure \sampling.} Lower panel : A histogram of the time
separations of quintuplets of consecutive measured points in the time
series of the gravitational lens QSO 0957+561. The upper panel shows
what this histogram would be if the quintuplets were uniformly spaced
over the same time span as the largest quintuplet separation in the actual
time series. The middle panel shows what this histogram would be if
the quintuplet separation was designed to cover this time span in a
non-redundant fashion.
}
\endfigure\nfig

With these $N=366$ quintuplets in the measured time series
${1\over 2} N (N-1) = 66795$ distinct pairs of distinct quintuplets can
be formed. For each distinct quintuplet pair the time difference between
the middle points is calculated and then binned into 5 day bins thus
accounting for the fact that even partially overlapping high sampling
rate sections are useful. The result is depicted as a histogram of
number of pairs per time lag bin. It is clear that if there are many
overlapping quintuplets for a given time separation then a time lag
in that range can be well determined by the high sampling rate
quintuplets, and only poorly if there are very few overlapping
quintuplets. For an ideal sampling strategy, without any prior knowledge
of the actual time lag, the quintuplet pair separations should cover
uniformly the entire range of time lags of interest.

\fignam{\sampexk}{sampexk}
\beginfigure*{6}
\epsfysize=11.4cm
\epsfbox{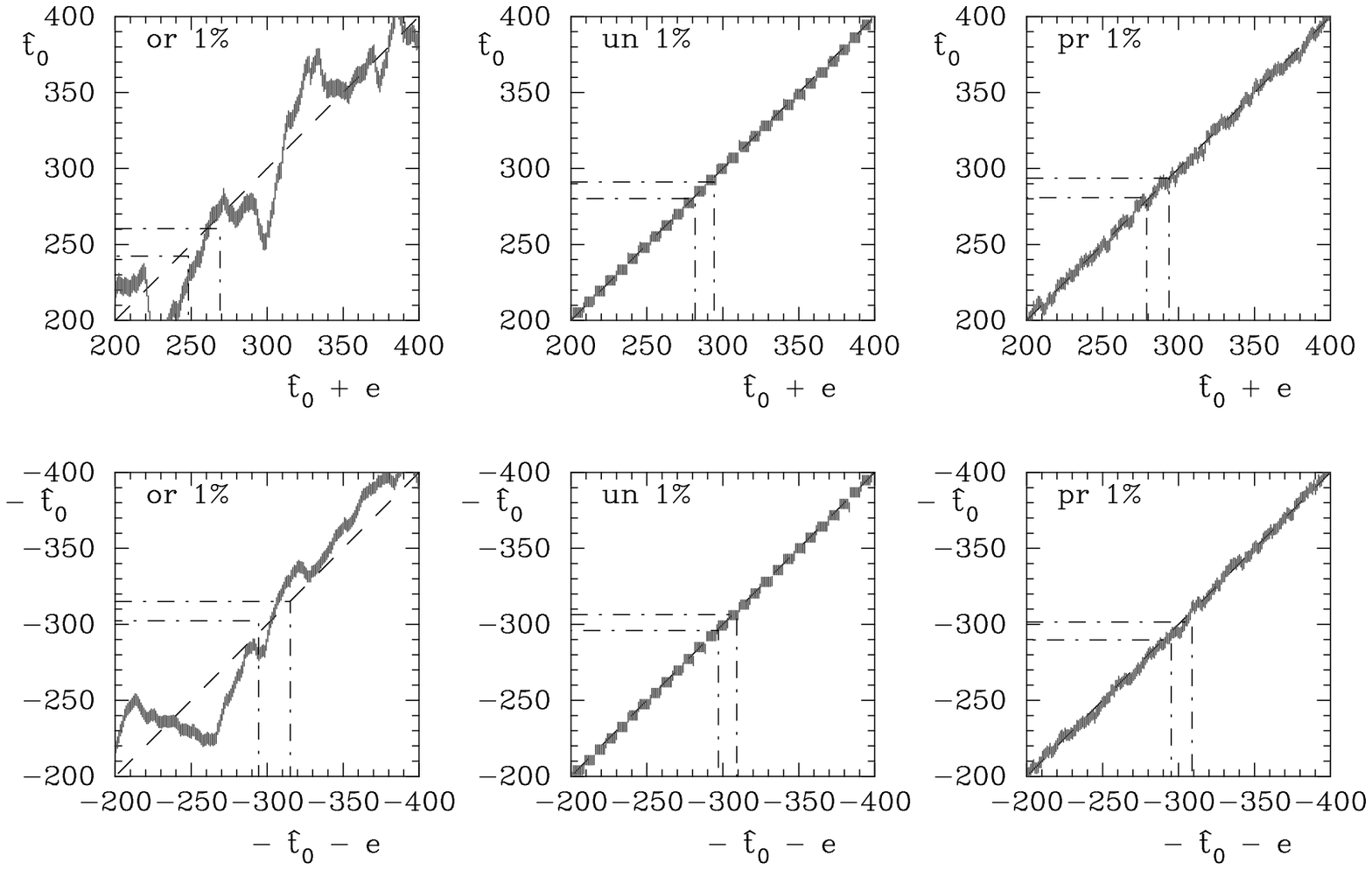}
\caption{{\bf Figure \sampexk.} The constructed kernels for the three
different sampling strategies of a simulated time series. The true delay
is $299\ {\rm d}$. The error weighting $\mu = 10^{-3}$ in all cases.
The $\tau_{\rm max}$ is $711$ for the original sampling of the data, $715$
for the unform sampling and $730$ for the prime number sampling. Only
the relevant subsection of the kernel is shown in order to demonstrate
the differences between kernels.}
\endfigure\nfig

In figure \sampling{} the resulting histogram is shown in the bottom
panel. The largest separation between two quintuplets in the measured time
series is $3514\ {\rm d}$ but the plotted histogram is restricted to
time lags between $0$ and $750\ {\rm d}$ which covers the range of
interest since the method presented here cannot reliably recover time lags
outside of this interval for the time series measured for QSO 0957+561.
For comparison in the top and middle panels two artificially designed
sampling strategies are shown. In the top panel the same number of $366$
quintuplets have been distributed uniformly over the time period covering
$3514\ {\rm d}$. It can be seen from figure \sampling{} that this results
in a bimodal pattern of separations. For some time lags no quintuplet
overlaps with another and for some lags quite a large number overlaps. The
few intermediate bins occur because the separation between successive
pairs is not an integer multiple of the bin width of $5\ {\rm d}$.
The second strategy shown in the middle panel of figure \sampling{} does
rather better. Here the $N = 366$ quintuplets are distributed over the
same time period covering $3514\ {\rm d}$ by putting them at the times
$T_5 (i)$ obtained by taking~:
\eqnam\prnumsamp{prnumsamp}
$$
T_5 (i) = { P_i P_{N+1} \over P_{2N+1-i} P_N }\times 3514\ {\rm d}\ \
\ \ \ \forall\ \ i=1,..,N
\eqno\neqn
$$
where the $P_i$ are prime numbers. Of course the resulting measurement times do
not fall on integer dates, so each measurement time is rounded to the
nearest integer before the separations are calculated and binned. This
strategy results in a much more nearly uniform coverage of the range of
possible time lags. The actual distribution of quintuplet pair separations
shown in the bottom panel of figure \sampling{} is clearly not uniform.
There are `gaps' around $150\ {\rm d}$ and between $200\ {\rm d}$ and
$250\ {\rm d}$ and between $470\ {\rm d}$ and $530\ {\rm d}$. Annual
occultation of the source by the Sun causes large gaps in the time series
which come back every year. This can diminish the number of available
quintuplet separations between time lags of around the time corresponding to
the length of that gap and around 1 yr minus that length. Similar dips
in the distribution of quintuplets separations would occur at the delays
of these same times with integer numbers of years added. However one can
compensate partially for this effect on the distribution of separations
by placing the quintuplets more often at the separations that currently
seem to be neglected. For a sun constrained measurement gap that lasts
less than half a year this is always possible, weather and telescope
scheduling permitting.

If the true time lag is between $350\ {\rm d}$ and $450\ {\rm d}$ then
the actual sampling of the time series appears fortunate, since in this
range there are very many pairs that overlap.
If the true time lag is between $450\ {\rm d}$ and $550\ {\rm d}$
however, the number of overlapping quintuplet pairs decreases to well
below what it would be in the nearly uniform case shown in the middle
panel of figure \sampling{}. Since there are claims in the literature
(cf. Press et al., 1992a,b) that the true time lag lies in this range
the sampling of this time series seems rather unfortunate, since
the time series is not best suited to test the longer lags. However,
one must keep in mind that this is only true in so far as the very high
sampling rate episodes are concerned. The overall time series can of
course be used for time lag determinations. The distribution of the high
sampling rate sections does demonstrate that the variations seen in those
sections can in general only with difficulty be treated as signal since
their counterparts in the time series for the other image are in
general not very likely to have been measured as well.

One should not conclude from this discussion that SOLA or any other
method is unable to recover a time delay if the sampling times are not
distributed uniformly or according to \prnumsamp). The simulated light curve of
section 3 did not follow this sampling and it is clear that a time delay
can be recovered. However the most accurate recovery for a given number
of samplings and a given photometric accuracy is obtained when the
separations in time between each pair of measurement points are designed
to be non-redundant. Weather conditions will always prevent the use of
a pre-designed strategy but by monitoring the measurement point-pair
separations while monitoring and scheduling subsequent measurements
to compensate for any dips in the distribution function of these
separations it should be possible to obtain a much more nearly
uniform distribution. In the following section the differences between
the various strategies is demonstrated using simulated time series.

\subsection{Comparing the results for different sampling strategies}

To demonstrate the effect of sampling strategies three sampling strategies
are used for the same simulated time series. For this simulation the time
series for image A is interpolated linearly and resampled uniformly, and
also resampled according to \prnumsamp). The second time series was obtained
by shifting the original by 511 days. Finally on all these six time series
random numbers drawn from a Gaussian distribution with zero mean and a standard
deviation of 1\% of the flux were added to mimic measurement errors.
This simulation is somewhat more realistic than the first example discussed
in section 3 because there is more high frequency signal in the image
A (and B) time series.

An inversion is carried out using the procedures outlined before.
In all cases the Savitzky-Golay fitting of the time series is done
with a window that is three points wide and a constant (0-order polynomial)
is fitted to these points. The error weighting $\mu = 10^{-3}$ in all cases.
The value of $\tau_{\rm max}$ is $711\ {\rm d}$ for the original sampling
of the data, $715\ {\rm d}$ for the uniform sampling and $730\ {\rm d}$
for the sampling according to \prnumsamp). The results are summarized
in table 2. The two artificial time lags were chosen to demonstrate the
difference between the accuracy for a time lag that should be well
sampled even with the original sampling of the A image, and a time lag
that should be more difficult to recover. The constructed linear kernels
for the case where the true $T_0 = 299\ {\rm d}$ are shown in figure
\sampexk{}. The results in table 2 show that there is negligible difference
between the uniform and prime number sampling results. From the
figure \sampexk{} it is clear however that the prime number strategy
produces a more uniform kernel. In the kernel for the regularly sampled
time series the bimodal pattern of quintuplet separations is reflected in
the regular block structure of the error bars on the kernel. An encouraging
result is that even for the original sampling the true time lag is quite
well recovered. In fact for this realization of the artificial errors
the recovery is marginally better for the longer time lag. More worrisome
is that the asymmetry in the results when interchanging the role of the time
series is quite large for the true sampling. Since the asymmetry is used
to determine possible contamination of the time series by extraneous effects
one might mistakenly conclude that the time series is contaminated.
Subsequently correcting for this non-existent contamination will produce
a time lag that is rather less accurate than the formal errors indicate.
It is in particular in the differences $\delta I$ and $Z$ that one
can see that a time lag inversion in the region of $500\ {\rm d}$
is more problematic than one in the region of $300\ {\rm d}$.

\begintable{2}
\caption{{\bf Table 2.} The results from the time lag inversions for
the artificial time series using different sampling strategies. The
entries in brackets are the propagated data errors which always apply
to the last decimal place of the entry to the left of it. The results
for two different artificial lags are shown. The columns show in order
the mean of the two determinations of $I$, and half the difference $\delta I$
between them, and $T_0$, and $Z$.}
\halign{\hfil#\hfil&\ \hfil#&#\ &\ \hfil#&#\ &\ \hfil#&#\ &\ \hfil#&#\cr
\noalign{\hrule\smallskip}
case & $I$ && $\delta I$ && $T_0$ && $Z$&\cr
\noalign{\smallskip\hrule\smallskip}
true & 1.0000 && 0.0000 && 299 && 0 &\cr
original & 0.9927 &(4)& 0.0005 &(4)& 282 &(6)& 23 &(6)\cr
uniform  & 0.9990 &(3)& 0.0003 &(3)& 295 &(4)&  8 &(6)\cr
prime no.& 0.9988 &(3)& 0.0001 &(3)& 294 &(5)&  8 &(6)\cr
\noalign{\smallskip\hrule\smallskip}
true & 1.0000 && 0.0000 && 511 && 0 &\cr
original & 0.9940 &(4)& 0.0023 &(4)& 514 &(6)& 66 &(6)\cr
uniform  & 0.9990 &(3)& 0.0002 &(3)& 515 &(4)&  7 &(6)\cr
prime no.& 0.9990 &(3)& 0.0003 &(3)& 511 &(5)&  9 &(6)\cr
\noalign{\medskip\hrule}}
\endtable

\section{The time lag for QSO 0957+516A,B}

\subsection{ The Schild \& Thomson data}

SOLA time lag determinations are carried out using various parameter
settings and with the time series obtained for the gravitational lens
QSO 0957+561 recently reported by Schild and Thomson (1995).
Parameters for the inversion are~:
\item{ - } $\tau_{\rm max}$. \par
\item{ - } The number of points in the moving window to interpolate the
time series under the integral sign $N_{\rm win}$. \par
\item{ - } The degree of the polynomial to fit to these points
$N_{\rm pol}$. \par
\item{ - } The error weighting parameter $\mu$. \par\noindent
The parameter settings for 3 cases are summarized in table 3.
The output of the SOLA method would then be the zero order moment $I$,
and the asymmetry $\delta I$ in this when interchanging the role
of the time series, and further $T_0$ and $Z$. Using the method described
in the appendix to correct for contamination, $\delta I$ and $Z$ are
replaced by a relative offset $y_1 \equiv C^{(a)} - C^{(b)}/I$ and a
relative drift $y_2 \equiv L^{(a)} - L^{(b)}/I$. The output is thus
four numbers~:\par
\item{ - } The relative offset $y_1$.\par
\item{ - } The relative drift $y_2$, assumed to be linear in time.\par
\item{ - } The relative magnification $I$.\par
\item{ - } The time-lag $T_0$.\par\noindent
The results for the three cases considered here are summarized in table 4.
Inversions with other parameter settings have been carried out. The
sensitivity of the errors to the error weighting is small and no systematic
shifts in the results have been found. The value of $\tau_{\rm max}$ is
quite strongly constrained by the time lag itself and by the total
length of the measured time series. Other values did not yield significantly
different results and usually had larger associated error estimates.
The cases shown here can thus be considered representative of the
most reliable determinations done with this method and these data.

\fignam{\detren}{detren}
\beginfigure{7}
\epsfxsize=7.5cm
\epsfbox{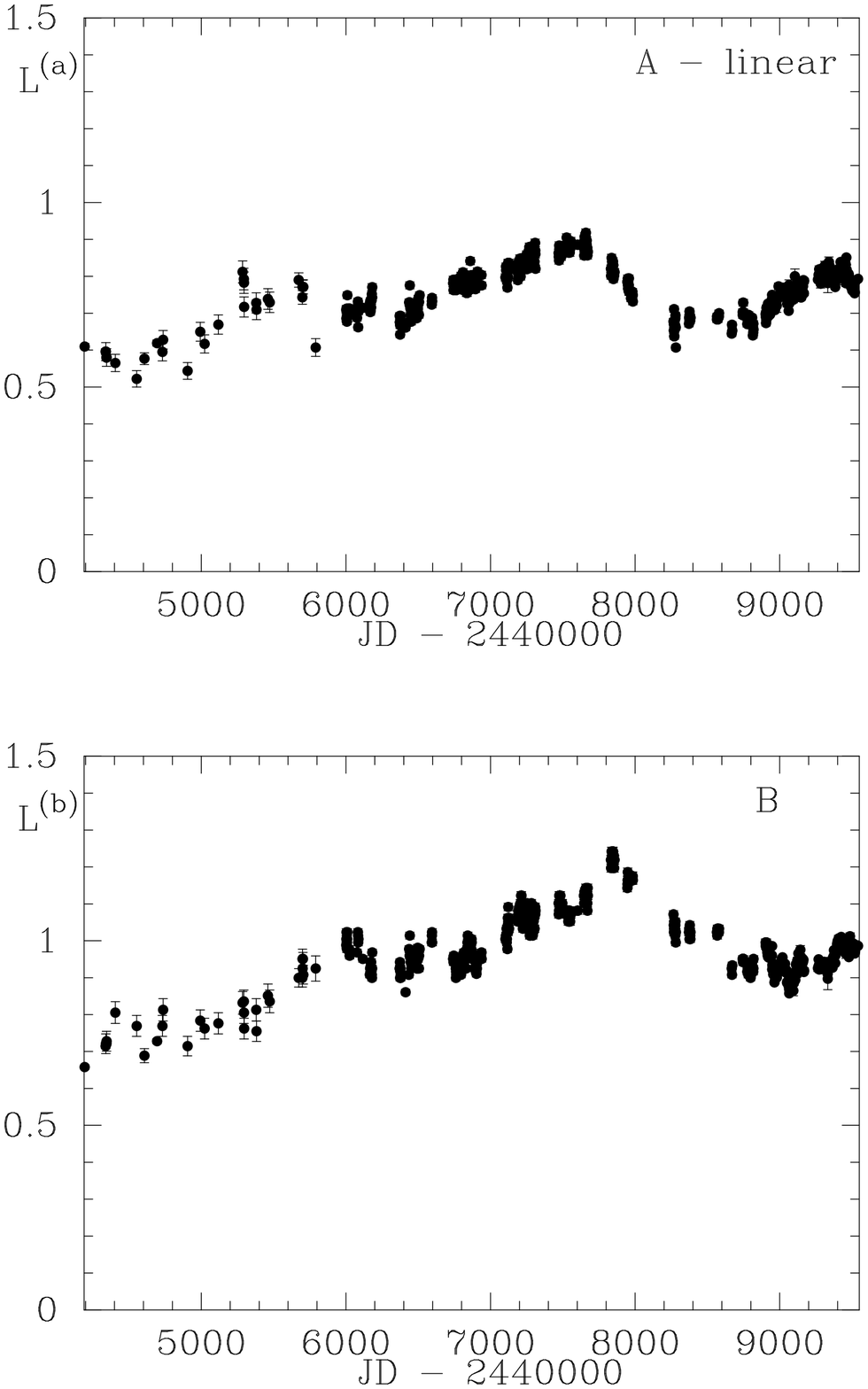}
\caption{{\bf Figure \detren.} The time series for image A and B of
the gravitational lens QSO 0957+561 after removing a linear function
of time expressing the offset and linear drift due to extraneous sources
(see the appendix). Only the difference in contamination between A and B can be
determined so the contamination is arbitrarily subtracted from only
the A image time series. }
\endfigure\nfig

\begintable*{3}
\caption{{\bf Table 3.} The various parameter settings in the time lag
inversions. For case 5 all points in quintuplets of daily sampling have
been merged as described in the text.}
\halign{\hfil#\hfil&\ \hfil #\hfil&\ \hfil# &\ \hfil#\hfil&\ \hfil#\hfil
&\ \hfil#\hfil&\ \hfil#\hfil&\ \hfil#\hfil&\ \hfil#\hfil\cr
\noalign{\hrule\smallskip}
case & $\tau_{\rm max}$ & $N_{\rm win}$ & $N_{\rm pol}$ & $\mu$ & $y_1$
& $y_2$ & $I$ & $T_0$ \cr
 & & & & & & $[\times 10^{-3}\ {\rm yr}^{-1}]$ & & [d] \cr
\noalign{\smallskip\hrule\smallskip}
 1 & 711 & 2 & 1 & $10^{-3}$ & $0.28 \pm 0.02$ & $-5.8 \pm 0.3$
& $1.34 \pm 0.06$ & $472 \pm 73$\cr
 2 & 711 & 3 & 0 & $10^{-3}$ & $0.29 \pm 0.01$ & $-6.8 \pm 0.2$
& $1.35 \pm 0.03$ & $425 \pm 17$\cr
 3 & 711 & 5 & 1 & $10^{-3}$ & $0.28 \pm 0.02$ & $-6.8 \pm 0.2$
& $1.30 \pm 0.05$ & $434 \pm 16$\cr
 4 & 711 &11 & 2 & $10^{-3}$ & $0.2  \pm 0.1$ & $-12 \pm 3$
& $1.2 \pm 0.3$ & $377 \pm 41$\cr
\noalign{\smallskip}
 5 & 711 & 3 & 0 & $10^{-3}$ & $0.21 \pm 0.02$ & $-6.8 \pm 0.2$
& $1.21 \pm 0.05$ & $469 \pm 16$\cr
\noalign{\medskip\hrule}}
\endtable

It is immediately clear from the asymmetries obtained in both the values
of $I$ and of the time lags $t_0$ when reversing the role of the time
series that some contamination of the time series is present. This
asymmetry is much larger than in the artificial time series with the
same sampling. The cause of the contamination can be found in e.g.
inaccurate subtraction of a foreground or background source, or
micro-lensing (cf. Schild and Smith 1991, Falco et al. 1991b). Following
the procedure outlined in the appendix the asymmetries are used to obtain
an estimate of this contamination. As described in the appendix it is
impossible to determine from this procedure which time series is
contaminated to what extent, since only a differential contamination
can be determined. The contamination in each time series is assumed to
be of the form $C + L (t-t_1)$.
In table 4 the offset $y_1 \equiv C^{(a)} - C^{(b)}/I$ and the drift
$y_2 \equiv L^{(a)} - L^{(b)}/I$. The flux units for $y_1$ and $y_2$ are
the same arbitrary units as used in equation \fluxun). Note that now
the asymmetries $\delta I$ and $Z$ are used in effect to determine
$y_1$ and $y_2$ so that after the corrections the $\delta I$ and $Z$ are
equal to $0$ to within $1 \sigma$.

Assuming for convenience that this contamination is entirely contained
in the A image the appropriate contribution is subtracted from the time
series of image A. Adding a similar contribution to image B instead produces
identical results in the subsequent inversions. The resulting time series
for case 2 are plotted in figure \detren. After this subtraction the
inversion yields the relative magnification $I$ and the time delay $T_0$.
The uncertainties quoted in table 4 are $1\sigma$ error bars due to the
propagation of the measurement errors of the time series. However one
should note that the error cross-correlation coefficients for these four
numbers are all in excess of $0.8$. In terms of a $\chi^2$ measure the
iso-$\chi^2$ contours in the 4-dimensional solution space form
hyper-ellipsoids with their major axes not aligned with the coordinate
axes. Taking this into account the actual uncertainty in for instance
the time lag is considerably larger, because an equal $\chi^2$ can be
achieved by simultaneously changing the other 3 values.
Furthermore one should note that the contamination is assumed to be
at most linear in time. As demonstrated by Schild and Smith (1991), and by
Falco et al. (1991b) there may well be higher-order terms present due
to the effect of micro-lensing.
While higher-order contamination will behave more as noise rather than
cause systematic shifts in the result, the quoted measurement error is
then no longer a good measure of the actual noise in the time series.
This effect also increases the uncertainty of the results.
The spread of points within quintuplets is roughly a factor of 2 times
the quoted error bars. If one assumes that the high frequency micro-lensing
events account for this excess spread one should increase the errors
by some amount, keeping in mind that such micro-lensing events do not
follow the same statistical distribution as measurement errors.

The contamination, the relative amplification factor and the time lag
shown in Table 4 for cases 1, 2 and 3 demonstrate that the inversions
are consistent to within the errors. This is what would be expected
for an ideal case where the errors are Gaussian and the distribution
in time of the sampling points is adequate for a time lag determination.
To assess the influence of possible high-frequency contaminating signal
from micro-lensing cases 4 and 5 were carried out. In case 4 the
interpolating window is set to 11 points to which a quadratic polynomial
is fitted. This should smooth out some of the variation within the window
but of course the detrimental effect of large gaps in the time series
are exacerbated by a wider window. An alternative approach is to isolate the
quintuplets of points measured on consecutive days and replace them
with their mean on the central day of the quintuplet. This is also a
smoothing operation but now carried out only on sections with a high
sampling rate. This is done in case 5. Both for case 4 and 5 it is
important to realize that this stronger smoothing is not specific. True
high frequency variations that are intrinsic to the lensed quasar
are removed together with the micro-lensing variations.

As a result of the smoothing one expects that the offset $y_1$ becomes
smaller in absolute value. The relative magnification factor is affected
as well however. This can be understood by writing the smoothing in terms
of an extra convolution with a smoothing function $S(t)$ that is
determined by the choice of window and fitting polynomial. Instead of
having a convolution as in equation \RevMap)~:
$$
L(t) = \Psi(t) * C(t)
\eqno\neqn$$
the convolution is now~:
$$
L(t) = \left( \Psi(t) * S^{-1}(t) \right) * \left( S(t) * C(t) \right)
\eqno\neqn$$
(or the analogous forms for \PhaseL) ). Here $S^{-1}(t)$ is the inverse
of the smoothing operation $S$. That is to say that because
the fit $S(t) * C(t)$ is used the transfer function $\Psi$ is modified
to $\Psi(t) * S^{-1}(t)$. When this is used in \PhaseL) and the steps
outlined before are followed to obtain the relative magnification $I$
it is clear that what is actually obtained is either $I S^{-1}(t_0)$ or
$I / S^{-1}(t_0)$. Increasing the smoothing leads to an increasing difference
between these two values. This increased asymmetry leads to an
overestimate of $y_1$ and an $I$ that will tend towards $1$. In the
limit where all variations in the time series are smoothed out one
cannot distinguish between the cases of time series that are offset due
to contamination and of time series that have a non-unity ratio between
the flux levels, which is the ratio between the average flux in either
series. The time lag and the drift $y_2$ should be less affected by
the smoothing as long as the smoothing window is symmetric around the
point at which the interpolation is evaluated. The error in $y_2$ and
$T_0$ should increase however because the accuracy with which a lag
can be determined depends in part on the intrinsic high frequency
signal. The results in Table 4 bear out the expectations in that not
only $y_1$ but also $I$ appear reduced. Because the $y_1$ is probably
overestimated the reduction of the contamination by the smoothing is
actually larger than the result shows because it is partially negated.
Considering the errors the various effects remain marginal however.
It is not clear whether the contamination is entirely due to
high frequency variations. The drift $y_2$ appears to be still present
and the effects of the smoothing should play only a minor role here.
The time lag for cases 4 and 5 are consistent
with the best determinations but the error is clearly higher for case 4.
On the basis of these results it seems that the best determination
of the time lag is case 2. An error weighted mean of all 5 time lag
determinations yields $T_0 = 441 \pm 9\ {\rm d}$. Considering that
these five determinations are not independent because the same data is
used it seems preferable to take the best case $T_0 = 425 \pm 17\ {\rm d}$
as final result.

\subsection{Other data sets}

In the literature, there are reported a number of attempts at
determining the time lag from optical (Schild \& Cholfin 1986,
Schild 1990, Pelt et al. (1994, 1996) at R band and B band by
Florentin-Nielsen (1984), Vanderriest et al. (1989), Press et al. (1992a)
Beskin \& Onkyanskij (1992) as well as radio data (Roberts et al 1992.
Lehar et al 1989, 1992, Press et al 1992b).
There is even an attempt using UV data (Gondhalekar et al. 1986).
The B band monitoring produces data that is similar to that used here
although generally the time series are shorter. Considering only the B band
data there appear to be emerging two mutually exclusive time lags, that
are obtained from the various analyses.
Press et al. analysed a shorter optical time series for QSO 0957+561
(1992a) and also a radio time series (1992b) using a rigorous statistical
method developed by them (1992a). They obtain an estimate for the time delay
of $536^{+14}_{-12}\ {\rm d}$ as 95~\% confidence interval from the
optical data and their determination from the radio data is consistent
with this value. An analysis by Pelt and coworkers (1994) of the same data
yields $415 \pm 32 \ {\rm d}$ and their analysis of the radio data is
consistent with their value for the optical time series. The determination
reported here appears at face value to be consistent with the determinations
of Pelt and coworkers (1994, 1996), and not with that of Press et al.
Considering the fact that the SOLA method can only correct for offsets
and linear drifts and also that
the correlation coefficient between the $y_1$, $y_2$, $I$, and $T_0$ are
all large means that it is not justified to exclude the estimates
of Press et al. out of hand. However Thomson and Schild (1996) argue that
the presence of variations due to micro-lensing invalidates some of the
assumptions made by Press et al. (1992a,b) concerning the statistical
properties of the time series. Thomson and Schild (1996) discuss how the Press
et al.
method can be corrected for this effect and then obtain a result that
is consistent with the shorter time lag found previously by them.
Since the study presented here also shows evidence for contamination of
the time series, although no specific cause can be identified from the
SOLA method itself, there is some support for their hypothesis.

To compare with previous determinations the SOLA method is also applied
to the same data (Vanderriest et al., 1989) that the method of Press et
al. was applied to. This data set spans a period of a bit over 2900 days
which is on the short side for a reliable estimate of a time lag using
SOLA, as argued in section 3. Furthermore the errors are larger for this
data set than for the Schild and Thomson data which leads to a larger
uncertainty in the time delay as well. The same parameter settings are
used for the SOLA algorithm as in case 3 for the Schild and Thomson data
except that $\tau_{\rm max}$ is reduced to $601\ {\rm d}$. The resulting
relative magnification is $1.0 \pm 0.14$ and the time delay is
$429 \pm 49\ {\rm d}$. The asymmetries $\delta I$ and $Z$ are on the level
of the $1 \sigma$ errors.

Finally Schild and Thomson (1995) present a master set of data from various
sources, including their own. This data set spans about the same time period
that Schild \& Thomson's own data do, but the coverage is somewhat better.
The SOLA parameter settings are as in case 2 in the previous section.
The resulting relative magnification $I = 1.30 \pm 0.03$, and the time delay
$T_0 = 496 \pm 12\ {\rm d}$. The contamination offset is $y_1 = 0.26 \pm 0.01$
and the drift $y_2 = -6.4 \pm 0.2 \times 10^{-3}$.
Although this set contains more measurements than the others considered here
one should realize that these data are not homogeneous in quality. If the
contamination in the time series is due in part to imperfect data reduction
procedures, which differ between the various observers, then it is a very poor
approximation to assume that the contamination can be described in terms of
a simple offset and drift. Furthermore the merging process itself of combining
these different parts of the time series can introduce extra systematic
errors. Taking this into account it seems unsurprising to find a difference
of some $5-7 \sigma$ between this result and the result when using only
the data of Schild and Thomson (1995).

\section{Obtaining $H_0$ from the time lag}

{}From a careful analysis of the images of QSO 0957+561A, B combined
with a tracing of the light deflection through a model potential
to fit the positions and magnifications of the images it is possible
to determine the gravitational potential of the lensing object. With
this potential it is possible to determine the Hubble constant as
long as the lensed source is variable.
This is essentially a consequence of the light of the two images having
traversed a different path through this gravitational potential. By
measuring the time delay for a signal propagating at the speed of light
an apparent separation of the images can be related to a physical distance
which together with a measured redshift yields an estimate of the Hubble
constant. Further details can be found in the monograph `Gravitational Lenses'
(Scheider et al. 1994)

{}From modeling of the lensing system Falco et al. (1991a) quote for the
value of the Hubble constant~:
\eqnam\HubA{HubA}
$$
H_0\ =\ (90 \pm 10)\left({\sigma_v \over 390\ {\rm km\ s^{-1}}}\right)^2
\left( {\tau_{AB}\over 1\ {\rm yr}} \right)^{-1}
\eqno\neqn
$$
The units are ${\rm km\ s^{-1}\ Mpc^{-1}}$, and $\sigma_v$ is the
velocity-dispersion of the lensing galaxy. $\tau_{AB}$ is the time
lag between the time series of the images.
More recent modeling by Grogin and Narayan (1996) yields the result~:
\eqnam\HubB{HubB}
$$
H_0\ =\ (82.5^{+8.7}_{-5.6} )\left({\sigma_v \over 322\ {\rm km\ s^{-1}}}
\right)^2 \left( {\tau_{AB}\over 1.1\ {\rm yr}} \right)^{-1}
\eqno\neqn$$
Using the value $\sigma_v = 303 \pm 50$ obtained by Rhee (1991),
a time delay of $425 \pm 17\ {\rm d}$, and using formula \HubA) yields~:
$$
H_0\ =\ 47 \pm 16\ {\rm km\ s^{-1}\ Mpc^{-1}}
\eqno\neqn
$$
Using instead \HubB) yields~:
$$
H_0\ =\ 69 \pm 24\ {\rm km\ s^{-1}\ Mpc^{-1}}
\eqno\neqn$$
The largest contribution to the error in these determinations is due to
the velocity dispersion. Grogin and Narayan (1996) also quote an equation
similar
to \HubB) for an upper limit to the Hubble constant which is independent
of the velocity dispersion~:
\eqnam\HubUp{HubUp}
$$
H_0\ =\ (82.5^{+8.7}_{-5.6} )\left( 1-\kappa \right)
\left( {\tau_{AB}\over 1.1\ {\rm yr}} \right)^{-1}\ {\rm km\ s^{-1}\ Mpc^{-1}}
\eqno\neqn$$
where $\kappa$ is the convergence of the quadratic potential with which
the cluster surrounding the lensing galaxy is approximated. Since
$\kappa$ must be positive, using the time delay yields an upper limit~:
$$
H_0 < 78 \pm 7\ {\rm km\ s^{-1}\ Mpc^{-1}}
\eqno\neqn$$
As Grogin and Narayan (1996) discuss the $\chi^2$ of their model for
the gravitational potential is somewhat lower than that of Falco et al.
(1991a) when using the most recent VLBI data as constraints for the model
parameters. For this reason equations \HubUp) and \HubB) are probably
more reliable than the older values of Falco et al. (1991a).

\section{Conclusions}

The SOLA method for inversion is modified to determine the time lag
between two time series, assuming that the width of the transfer
function between the two is small compared to the typical sampling time
interval. The method is demonstrated to perform well on artificial data.
The method is then applied to the measured time series of the
gravitational lens QSO 0957+561. Considering the uncertainties in the
de-trending the best estimate obtained here gives a value for the
time delay of $425 \pm 17\ {\rm d}$. This leads to a best value for
the Hubble constant of $H_0\ =\ 69 \pm 24\ {\rm km\ s^{-1}\ Mpc^{-1}}$.

Considering the primary sources of uncertainty in the determination of
the Hubble constant from the time delay for QSO 0957+516 it appears
highly desirable to obtain a more accurate determination for the velocity
dispersion of the lensing system. The determination of the time lag itself
could be improved upon substantially if an independent means can be found to
determine quantitatively the effects that contaminate the time series
such as micro-lensing but also the effects of merging data from various
sources and observers. The SOLA time lag determinations from Schild \&
Thomson's own data, and from their `masterset' differ by as much as $5-7$
times the error propagated from the errors in photometry. It seems that
there are systematic errors in at least one of these two data sets that
have not yet been accounted for. Presumably this would affect any method
to determine time delays and so one should not be surprised to find
a spread in time delay results in the literature that is larger than
the formal errors quoted.

\section*{Acknowledgments}
The author thanks V. Oknyanskij and E. van Groningen for pointing out
this problem. R. Schild is thanked for sending the time series of
QSO~0957+561A, B in electronic form, and for pointing out various
factors contaminating the time series. I. Wanders is thanked for many
helpful suggestions and discussions regarding the testing of the algorithm.

\section*{Appendix : Contamination of the time series}

The analysis in the main paper is based on the assumption that the two time
series can be measured without any contamination by foreground sources or
instrumental offsets or drifts, so that only random (measurement) noise
is a source of errors. If one or both images of the quasar are contaminated
by a foreground source, for instance the lensing object itself, this may
influence the time lag determination.
If the foreground source is rapidly varying it can only give rise to false lags
or aliases if that variability is significantly correlated with the time
series of the lensed quasar. This is a priori unlikely because there is no
causal physical connection between the two objects. In practice it may occur
if the measured time series is short compared to the time scales of variability
of the lensed quasar and the hypothetical contaminating foreground source or
if the physical mechanisms causing the variability in the lens and the quasar
have matching characteristic time scales. In the absence of such
correlation the variable part merely behaves as an extra source of noise
in the time series which can increase the uncertainty in the determined lag,
but it will not have a systematic effect.
A constant or low frequency non-zero contamination in one or both images
can influence the determination of the time lag indirectly. It is
demonstrated in this appendix that this can be detected if it occurs and
can be corrected for.

Consider again equations \PhaseL). Now, instead of having measured $F^{(a)}$
and $F^{(b)}$, contaminated time series $\topTwi{F}^{(a)}$ and
$\topTwi{F}^{(b)}$ are measured~:
\eqnamA\contamPL{contamPL}
$$
\cases{
\topTwi{F}_i^{(a)}\ \equiv F_i^{(a)} + C^{(a)} + L^{(a)} (t_i-t_1)\cr
\topTwi{F}_i^{(b)}\ \equiv F_i^{(b)} + C^{(b)} + L^{(b)} (t_i-t_1)\cr
}
\eqno\neqnA
$$
in which $C^{(a)}$ and $C^{(b)}$ are the (unknown) constant contaminating
source(s) and the terms with coefficients $L$ represent a possible (linear)
drift.
With these contaminated time series the kernels are built, and the inversion
is carried out. In the main paper the normalization of the base functions
is taken implicit for convenience of notation. Here these factors
are be kept separate as coefficients $\{u_i\}$ or, for the case of the
contaminated series, $\{\toptwi{u}_i\}$~:
$$
\eqalign{
\toptwi{u}_i^{(a)}\ &\equiv\ \left(\int\limits_{-\tau_{\rm max}}^{
\tau_{\rm max}} \topTwi{F}^{(a)} (t_i +\tau) {\rm d} \tau\right)^{-1}  \cr
\toptwi{u}_i^{(b)}\ &\equiv\ \left(\int\limits_{-\tau_{\rm max}}^{
\tau_{\rm max}} \topTwi{F}^{(b)} (t_i +\tau) {\rm d} \tau\right)^{-1}  \cr}
\eqno\neqnA
$$
These normalization factors are calculated at the start of the algorithm
and hence are known. For convenience of notation the following
linear combinations $g$ and $h$ are defined~:
$$
\eqalign{
g^{(0a)}\ &\equiv\ \sum \toptwi{c}_i^{(0a)} \toptwi{u}_i^{(a)} \quad,\quad
g^{(1a)}\ \equiv\ \sum \toptwi{c}_i^{(1a)} \toptwi{u}_i^{(a)} \cr
g^{(0b)}\ &\equiv\ \sum \toptwi{c}_i^{(0b)} \toptwi{u}_i^{(b)} \quad,\quad
g^{(1b)}\ \equiv\ \sum \toptwi{c}_i^{(1b)} \toptwi{u}_i^{(b)} \cr}
\eqno\neqnA
$$
$$
\eqalign{
h^{(0a)}\ &\equiv\ \sum \toptwi{c}_i^{(0a)} \toptwi{u}_i^{(a)} (t_i-t_1) \cr
h^{(1a)}\ &\equiv\ \sum \toptwi{c}_i^{(1a)} \toptwi{u}_i^{(a)} (t_i-t_1) \cr
h^{(0b)}\ &\equiv\ \sum \toptwi{c}_i^{(0b)} \toptwi{u}_i^{(b)} (t_i-t_1) \cr
h^{(1b)}\ &\equiv\ \sum \toptwi{c}_i^{(1b)} \toptwi{u}_i^{(b)} (t_i-t_1) \cr}
\eqno\neqnA
$$
For the zero-order moment the following holds~:
\eqnamA\contCombZ{contCombZ}
$$
\eqalign{
\sum \toptwi{c}_i^{(0a)} \toptwi{u}_i^{(a)} \topTwi{F}_i^{(a)} &=
{1\over 2\tau_{\rm max}} \cr
\sum \toptwi{c}_i^{(0a)} &= 1\cr}
\eqno\neqnA
$$
Combining \contCombZ) with \contamPL) yields~:
$$
\eqalign{
\sum \toptwi{c}_i^{(0a)} \toptwi{u}_i^{(a)} F_i^{(a)} &= {1\over
2\tau_{\rm max}} - C^{(a)} g^{(0a)} - L^{(a)} h^{(0a)} \cr
\sum \toptwi{c}_i^{(0a)} \toptwi{u}_i^{(a)} \topTwi{F}_i^{(b)} &=
\sum \toptwi{c}_i^{(0a)} \toptwi{u}_i^{(a)} F_i^{(b)} + C^{(b)} g^{(0a)} \cr
&\qquad\qquad + L^{(b)} h^{(0a)} \cr}
\eqno\neqnA
$$
Working through the equivalent of equations \steps) it can be seen that
instead of determining the constant $I$ what is determined with the
contaminated series is~:
\eqnamA\Itwids{Itwids}
$$
\eqalign{
\topTwi{I}^{(a)} = I + 2 \tau_{\rm max} g^{(0a)}&\left[ C^{(b)} -
I C^{(a)} \right] \cr
&+ 2 \tau_{\rm max} h^{(0a)}\left[ L^{(b)} - I L^{(a)} \right] \cr
{1\over\topTwi{I}^{(b)}} = {1\over I} + 2 \tau_{\rm max} g^{(0b)}
&\left[ C^{(a)} - {1\over I} C^{(b)} \right] \cr
&+ 2 \tau_{\rm max} h^{(0b)}\left[ L^{(a)} - {1\over I} L^{(b)} \right] \cr}
\eqno\neqnA
$$
For the linear target kernel ${\cal T}\ = \tau$ the coefficients
$\toptwi{c}^{(1a)}$ satisfy~:
\eqnamA\contComb{contComb}
$$
\eqalign{
\sum \toptwi{c}_i^{(1a)}\toptwi{u}_i^{(a)} \topTwi{F}_i^{(a)} &= \tau \cr
\sum \toptwi{c}_i^{(1a)} &= 0 \cr}
\eqno\neqnA
$$
Combining \contComb) with \contamPL) yields~:
$$
\eqalign{
\sum \toptwi{c}_i^{(1a)} \toptwi{u}_i^{(a)} F_i^{(a)} &= \tau - C^{(a)}
g^{(1a)} - L^{(a)} h^{(1a)} \cr
\sum \toptwi{c}_i^{(1a)} \toptwi{u}_i^{(a)} \topTwi{F}_i^{(b)} &= \sum
\toptwi{c}_i^{(1a)} \toptwi{u}_i^{(a)} F_i^{(b)} + C^{(b)} g^{(1a)} \cr
&\qquad\qquad + L^{(b)} h^{(1a)} \cr}
\eqno\neqnA
$$
The same holds of course when the role of the time series is reversed.
The resulting first order moments are~:
\eqnamA\ITtwids{ITtwids}
$$
\eqalign{
\topTwi{I t_0}^{(a)} = It_0 + g^{(1a)}&\left[ C^{(b)} - I C^{(a)} \right] \cr
&+ h^{(1a)}\left[ L^{(b)} - I L^{(a)} \right] \cr
{\topTwi{t_0}\over\topTwi{I}^{(b)}} = {t_0\over I} - g^{(1b)} &\left[
C^{(a)} - {1\over I} C^{(b)} \right] \cr
&- h^{(1a)}\left[ L^{(a)} - {1\over I} L^{(b)} \right] \cr}
\eqno\neqnA
$$
Clearly a systematic deviation is present due to the contaminating
source in both the zero order and first order moments. The presence of
noise in the measured time series means that the equality in the first of
equations \contCombZ) and of \contComb) become approximate just as in
equations \conskern). This does not affect the determination in any
systematic way.

The equations \Itwids) and \ITtwids) form a system of four equations in
the four unknowns $I$, $t_0$, $y_1 \equiv C^{(a)} - C^{(b)}/I$, and
$y_2 \equiv L^{(a)} - L^{(b)}/I$. Note that taking $y_1$ and $y_2$
in this way reflects that with this method it is impossible to tell
which of the two time series are contaminated, or to what extent either
of them is contaminated.

Since this system of equations is non-linear it is possible that no
solution exists or more than one solution. In any case the propagation
of the data errors could cause large uncertainties in the parameters.
Therefore the procedure that is followed is to solve \Itwids) and \ITtwids)
to determine $y_1$ and $y_2$. The appropriate contamination is then
subtracted from one of the two time series and the inversion procedure
is carried out again.
For data free of random errors this second iteration should be
contamination-free. For real data a few iterations can be necessary to
reduce the contamination to zero within the measurement
errors and an accurate determination of the relative magnification factor
$I$ and the time delay $t_0$ is then possible directly from the inversion
as outlined in the main part of this paper.

\bye